\documentclass[preprint,prb]{revtex4}%
\usepackage{amsfonts}
\usepackage{amsmath}
\usepackage{amssymb}
\usepackage{graphicx}%
\begin{document}
\title{Compressibility of Ce$M$In$_{5}$ and Ce$_{2}M$In$_{8}$ ($M=$Rh, Ir and Co) Compounds}
\author{Ravhi S. Kumar and A.L. Cornelius}
\affiliation{Department of Physics, University of Nevada, Las Vegas, Nevada, 89154-4002}
\author{J.L. Sarrao}
\affiliation{Materials Science and Technology Division, Los Alamos National Laboratory, Los
Alamos, NM\ 87545}

\begin{abstract}
The lattice parameters of the tetragonal compounds Ce$M$In$_{5}$ and Ce$_{2}%
M$In$_{8}$ ($M=$Rh, Ir and Co) have been studied as a function of pressure up
to 15 GPa using a diamond anvil cell under both hydrostatic and
quasihydrostatic conditions at room temperature. The addition of $M$In$_{2}$
layers to the parent CeIn$_{3}$ compound is found to stiffen the lattice as
the 2-layer systems (average of bulk modulus values $B_{0}$ is 70.4 GPa) have
a larger $B_{0}$ than CeIn$_{3}$ (67 GPa), while the 1-layer systems with the
are even stiffer (average of $B_{0}$ is 81.4 GPa). Estimating the
hybridization using parameters from tight binding calculations shows that the
dominant hybridization is $fp$ in nature between the Ce and In atoms. The
values of $V_{pf}$ at the pressure where the superconducting transition
temperature $T_{c}$ reaches a maximum is the same for all Ce$M$In$_{5}$
compounds. By plotting the maximum values of the superconducting transition
temperature $T_{c}$ versus $c/a$ for the studied compounds and Pu-based
superconductors, we find a universal $T_{c}$ versus $c/a$ behavior when these
quantities are normalized appropriately. These results are consistent with
magnetically mediated superconductivity.

\end{abstract}
\date{Today}
\maketitle

\section{Introduction}

Ce based heavy fermion (HF) and antiferromagnetic (AF) compounds have been the
subject of intensive investigations due to their unconventional magnetic and
superconducting properties. In these compounds the electronic correlations,
the magnetic ordering temperature and the crystal field effects are sensitive
to pressure, and pressure induced superconductivity near a quantum critical
point (QCP) has been observed in a variety of compounds such as CePd$_{2}%
$Si$_{2}$, CeCu$_{2}$Ge$_{2}$, CeRh$_{2}$Si$_{2}$ and CeIn$_{3}$.
\cite{Steglich79,Jaccard92,Movshovich96,Grosche96,Mathur98} The appearance of
superconductivity in these systems and the deviation from Fermi liquid
behavior as a function of pressure are still challenging problems to be studied.

Ce$_{n}M$In$_{2n+3}$ ($M$=Rh, Ir and Co) with $n=1$ or 2 crystallize in the
quasi-two-dimensional (quasi-2D) tetragonal structures Ho$_{n}$CoGa$_{2n+3}%
.$\cite{Grin79,Grin86} The crystal structure can be viewed as (CeIn$_{3}%
$)$_{n}$($M$In$_{2}$) with alternating $n$ (CeIn$_{3}$) and ($M$In$_{2}$)
layers stacked along the $c$-axis. By looking at the crystal structure, we
would expect that AF correlations would develop in the cubic (CeIn$_{3}$)
layers in a manner similar to bulk CeIn$_{3}$.\cite{Lawrence80} The AF
(CeIn$_{3}$) layers will then be weakly coupled by an interlayer exchange
interaction through the ($M$In2) layer leading to a quasi-2D magnetic
structure. Indeed, in the Rh compounds, the magnetic properties, as determined
by thermodynamic,\cite{Cornelius00} NQR,\cite{Curro00} and neutron scattering
\cite{Bao00} are less 2D as the crystal structure becomes less 2D going from
single layer CeRhIn$_{5}$ to double layer Ce$_{2}$RhIn$_{8}$ (note that as
$n\rightarrow\infty$, one gets the 3D cubic system CeIn$_{3}$). CeRhIn$_{5}$
and Ce$_{2}$RhIn$_{8}$ are antiferromagnets at ambient pressure but are found
to superconduct at high pressures.\cite{Hegger00,Fisher02,Mito01,Nicklas03}
The systems CeCoIn$_{5}$, CeIrIn$_{5}$ and Ce$_{2}$CoIn$_{8}$ display
superconductivity at ambient
pressure.\cite{Fisher02,Mito01,Petrovic01,Petrovic01_2,Chen02}. The only
member of the series that does not display magnetic order or superconductivity
at ambient pressures is Ce$_{2}$IrIn$_{8}$ that is believed to be near a
QCP.\cite{Thompson01}

While not proven definitively, it is generally believed that the origin of the
superconductivity in Ce$_{n}M$In$_{2n+3}$ is magnetic in origin. The value of
the superconducting transition temperature $T_{c}$ in magnetically mediated
superconductors is believed to be dependent on dimensionality in addition to
the characteristic spin fluctuation temperature. Theoretical models and
experimental results suggest that the SC state in CeRhIn$_{5}$ may be due to
the quasi-two dimensional (2D) structure and anisotropic AF fluctuations which
are responsible for the enhancement of\ $T_{c}$ relative to CeIn$_{3}%
$.\cite{Pagliuso02,Monthoux01} A\ strong correlation between the ambient
pressure ratio of the tetragonal lattice constants $c/a$ and $T_{c}$ in the
Ce$M$In$_{5}$ compounds is indicative of the enhancement of the
superconducting properties by lowering dimensionality (increasing $c/a$
increases $T_{c}$).\cite{Pagliuso02} In order to explain the evolution of
superconductivity induced by pressure and the suppression of AF ordering, it
is important to probe the effect of pressure on structure for these compounds
and look for possible correlations between structural and thermodynamic properties.

Here we report on high pressure x-ray diffraction measurements performed on
Ce$_{n}M$In$_{2n+3}$ ($M$=Rh, Ir and Co) with $n=1$ or 2 up to 15 GPa under
both hydrostatic and quasihydrostatic conditions. Previously, we have reported
results on CeRhIn$_{5}$;\cite{Kumar04} we present a comparative study of the
complete set of Ce$_{n}M$In$_{2n+3}$ compounds with emphasis on the behavior
near the QCP. While there is no direct correlation between $c/a(P)$ and
$T_{c}(P)$ as an implicit function of pressure in an individual system, the
value of $c/a$ at the pressure where $T_{c}$ reaches its maximum value DOES
show linear behavior as previously hypothesized.\cite{Pagliuso02} Also, the
$pf$ hybridization $V_{pf}$ between the Ce and In atoms is the dominant
hybridization in these compounds and takes on the same value for all
Ce$M$In$_{5}$ compounds at the pressure $P_{\max}$ where $T_{c}$ reaches its
maximum value. These results will be compared to isostructural Pu compounds
and all of the results are consistent with unconventional, magnetically
mediated superconductivity.

\section{ Experiment}

Ce$_{n}M$In$_{2n+3}$ single crystals were grown by a self flux technique
described elsewhere.\cite{Moshopoulou01} The single crystals were crushed into
powder and x-ray diffraction measurements show the single phase nature of the
compound. In agreement with previous results,\cite{Moshopoulou01,Moreno02} the
crystals were found to have tetragonal symmetry with cell parameters in
agreement with literature values.

The high pressure x-ray diffraction (XRD) experiments were performed using a
rotating anode x-ray generator (Rigaku) for the quasihydrostatic runs and
synchrotron x-rays at HPCAT, Sector 16 at the Advanced Photon Source for
hydrostatic measurements. The sample was loaded with NaCl or ruby powder as a
pressure calibrant and either a silicone oil or 4:1 methanol:ethanol mixture
(hydrostatic) or NaCl (quasihydrostatic) as the pressure transmitting medium
in a Re gasket with a $180$ $\mu$m\ diameter hole. High pressure was achieved
using a Merrill-Basset diamond anvil cell with $600$ $\mu$m culet diameters.
The XRD patterns are collected using an imaging plate ($300\times\ 300$
mm$^{2}$ ) camera with $100\times\ 100$ $\mu$m$^{2}$ pixel dimensions. XRD
patterns were collected up to 15 GPa at room ($T=295$ K) temperature. The
images were integrated using FIT2D software.\cite{Hammersley96} The structural
refinement of the patterns was carried out using the Rietveld method on
employing the FULLPROF and REITICA (LHPM)\ software
packages.\cite{Rodriguez-Carvajal93}

\section{ Results and Discussion}%

\begin{figure}
[ptb]
\begin{center}
\includegraphics[
height=3.4627in,
width=2.7769in
]%
{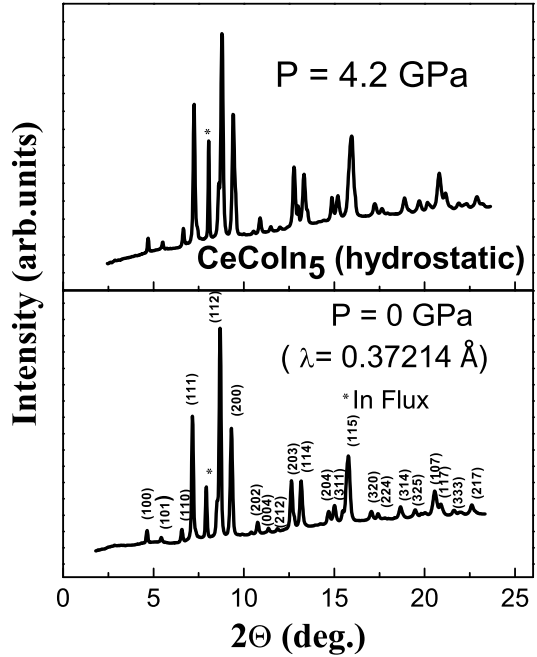}%
\caption{X-ray diffraction patterns of CeCoIn$_{5}$ at ambient pressure and a
hydrostatic pressure of 4.2 GPa. The data were taken using synchrotron
radiation of wavelength $\lambda=0.37214$ \AA . The various reflections from
CeCoIn$_{5}$ are labeled and one peak due to excess In flux is noted.}%
\label{xrd}%
\end{center}
\end{figure}
In Fig. \ref{xrd} we show the XRD patterns for CeCoIn$_{5}$ obtained at
ambient pressure and a hydrostatic pressures of 4.2 GPa with silicone oil used
as the pressure transmitting media. In other measurements, diffraction peaks
from the Re gasket, pressure markers (NaCl) and the sample are all observed.
The known equation of state for NaCl \cite{Brown99} or the standard ruby
fluorescence technique \cite{Piermarini75} was used to determine the pressure.
The refinement of the XRD patterns was performed on the basis of the Ho$_{n}%
$CoGa$_{2n+3}$ structure with the P4/mmm space group (No. 123). When comparing
the crystallographic data and bulk modulus of CeIn$_{3}$ relative to Ce$_{n}%
M$In$_{2n+3}$ it is evident that the Ce atom in Ce$_{n}M$In$_{2n+3}$
experiences a chemical pressure at ambient
conditions,\cite{Hegger00,Cornelius00} leading one to expect the Ce$_{n}%
M$In$_{2n+3}$ to be less compressible than CeIn$_{3}$ as the bulk modulus
increases with increasing pressure.%

\begin{figure}
[ptb]
\begin{center}
\includegraphics[
height=4.3716in,
width=1.5921in
]%
{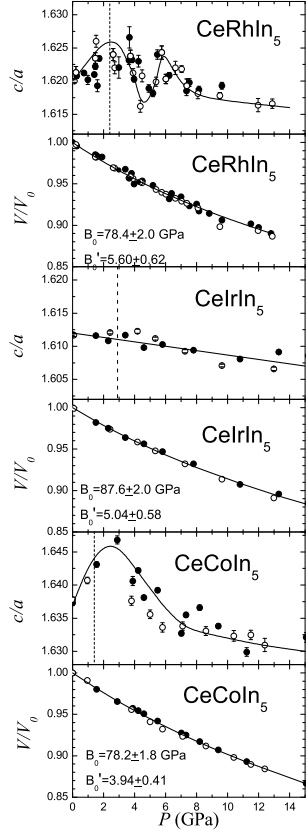}%
\caption{Normalized volume $V/V_{0}$ and ratio of tetragonal lattice constants
$c/a$ plotted versus pressure for Ce$M$In$_{5}$ compounds at room temperature.
Data for both quasihydrostatic (open symbols) and hydrostatic (closed symbols)
are displayed. The solid line through the volume data is a fit as described in
the text. The dashed vertical lines in the $c/a$ plots shows the pressure
where the maximum value of $T_{c}$ is observed. The solid lines in the $c/a$
plots are guides for the eye.}%
\label{cemin5}%
\end{center}
\end{figure}
\begin{figure}
[ptbptb]
\begin{center}
\includegraphics[
height=2.994in,
width=1.5887in
]%
{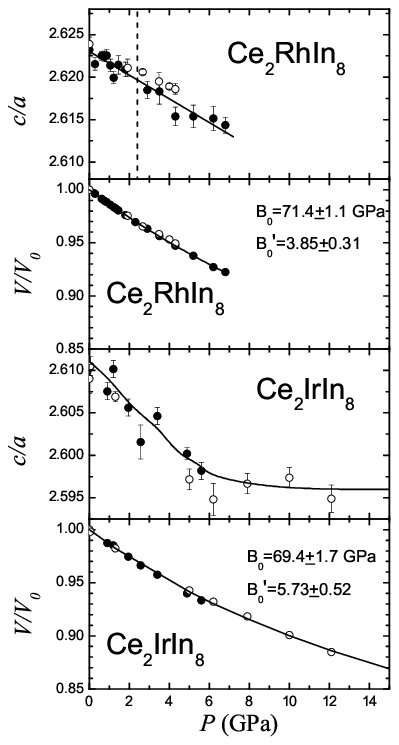}%
\caption{Normalized volume $V/V_{0}$ and ratio of tetragonal lattice constants
$c/a$ plotted versus pressure for Ce$_{2}M$In$_{8}$ compounds at room
temperature. Data for both quasihydrostatic (open symbols) and hydrostatic
(closed symbols) are displayed. The solid line through the volume data is a
fit as described in the text. The dashed vertical lines in the $c/a$ plots
shows the pressure where the maximum value of $T_{c}$ is observed. The solid
lines in the $c/a$ plots are guides for the eye.}%
\label{ce2min8}%
\end{center}
\end{figure}

The $V(P)$ data have been plotted in Fig. \ref{cemin5} for Ce$M$In$_{5}$
($M$=Rh, Ir and Co) and Fig. \ref{ce2min8} for Ce$_{2}M$In$_{8}$ ($M$=Rh or
Ir) for both quasihydrostatic and hydrostatic measurements (the data for
CeRhIn$_{5}$ has been previously reported\cite{Kumar04}). Note that the
vertical and horizontal scales are the same for all graphs.
Unfortunately, we have not had success growing single crystals of Ce$_{2}%
$CoIn$_{8}$, though others have reported successful growth of single
crystals.\cite{Chen02} Since the maximum volume compression is only of the
order of 10\%, the $V(P)$ data has been fit using a least squares fitting
procedure to the first order Murnaghan equation of state%
\begin{equation}
P=\frac{B_{0}}{B_{0}^{\prime}}\left[  \left(  \frac{V_{0}}{V(P)}\right)
^{B_{0}^{\prime}}-1\right]  ,
\end{equation}
where $B_{0}$ is the initial bulk modulus and $B_{0}^{\prime}$ is the pressure
derivative of $B_{0}$. For the room temperature ($T=295$ K) data $V/V_{0}$
data shown in Figs. 2 and 3, the values of $B_{0}$ and $B_{0}^{\prime}$ and
the initial linear compressibilities $\kappa_{a}$ and $\kappa_{c}$ calculated
below 2 GPa are given in Table \ref{boandboprime}.\begin{table}[ptbptbptb]
\narrowtext%
\begin{tabular}
[c]{|c|c|c|c|c|c|c|c|}\hline
System & $n$ & $V_{0}$(\AA $^{3}$) & $c/a$ & $B_{0}$ (GPa) & $B_{0}^{\prime}$
& $\kappa_{a}$(10$^{-3}$ GPa$^{-1}$) & $\kappa_{c}$(10$^{-3}$ GPa$^{-1}%
$)\\\hline
CeRhIn$_{5}$ & 1 & 163.03 & 1.621 & $78.4\pm2.0$ & $5.60\pm0.62$ &
$3.96\pm0.08$ & $4.22\pm0.10$\\\hline
CeIrIn$_{5}$ & 1 & 163.67 & 1.612 & $87.6\pm2.0$ & $5.04\pm0.58$ &
$3.44\pm0.06$ & $3.48\pm0.08$\\\hline
CeCoIn$_{5}$ & 1 & 160.96 & 1.638 & $78.2\pm1.8$ & $3.94\pm0.41$ &
$4.35\pm0.08$ & $3.43\pm0.16$\\\hline
Ce$_{2}$RhIn$_{8}$ & 2 & 266.48 & 2.624 & $71.4\pm1.1$ & $3.85\pm0.31$ &
$4.20\pm0.04$ & $4.85\pm0.11$\\\hline
Ce$_{2}$IrIn$_{8}$ & 2 & 266.26 & 2.610 & $69.4\pm1.7$ & $5.73\pm0.52$ &
$4.02\pm0.06$ & $4.93\pm0.12$\\\hline
CeIn$_{3}$ & $\infty$ & 103.10 & 1 & $67.0\pm3.0$ & $2.5\pm0.5$ &
$4.98\pm0.13$ & $4.98\pm0.13$\\\hline
\end{tabular}
.\caption{Summary of the determined bulk modulus $B_{0}$ and its pressure
derivative $B_{0}^{\prime}$ as determined from fits to the Murnaghan equation
for the Ce$_{n}M$In$_{2n+3}$ compounds. Also listed are the ambient pressure
values of $V_{0}$ and $c/a$ along with the initial linear compressibilities
$\kappa_{a}$ and $\kappa_{a}$. Values for CeIn$_{3}$ are taken from Vedel
\textit{et al}.\cite{Vedel87}}%
\label{boandboprime}%
\end{table} First, we note that the $n=2$ compounds show more anisotropy
($\kappa_{a}$ is 15-20\% smaller than $\kappa_{c}$) in the the
compressibilities than the $n=1$ compounds. As mentioned, the $n=1$ compounds
appear to be more 2D than the $n=2$ compounds, making this result somewhat
surprising. We also note the deviation from the typical inverse relationship
between $B_{0}$ and $V_{0}$; namely, CeIrIn$_{5}$ has the largest value of
$B_{0}$ AND\ the largest ambient pressure volume. These results hint that the
valence of Ce and hybridization between the Ce 4f electrons and the conduction
electrons needs to be taken into account. Pressure is known to make Ce
compounds more tetravalent, and since the tetravalent ion is smaller then the
trivalent ion, makes the more tetravalent system less compressible. The
explanation for the unexpected difference in the linear compressibilities may
lie in the fact that $c/a$ seems to be coupled to $T_{c}$ as will be discussed
later. As a larger $c/a$ favors superconductivity, if pressure reduces $c$
less than expected, the compressibility will be lowered and the $c/a$ ratio
will increase as seen in CeRhIn$_{5}$ and CeCoIn$_{5}.$ As expected, the
lattice appears to be stiffer the more 2D the system becomes as the $M$%
In$_{2}$ layers in Ce$_{2}M$In$_{8}$ stiffen the structure relative to
CeIn$_{3}$. CeIn$_{3}$ has a smaller bulk modulus ($B_{0}=67$
GPa)\cite{Vedel87} than the 2-layer systems (average of $B_{0}$ is 70.4 GPa)
that in turn is smaller than the 1-layer systems (average of $B_{0}$ is 81.4
GPa). The bulk modulus values compare well with those reported for other HF
systems \cite{Penney82,Spain86,Kutty87,Wassilew-Reul97}. The fact that we see
no discernible difference between the hydrostatic and quasihydrostatic
measurements is likely due to the nearly isotropic compressibilities.

Figs. 2 and 3 also show the ratio of the lattice constants $c/a$ as a function
of pressure. The systems display a wide range of behavior from the apparent
double peaked structure in CeRhIn$_{5}$ to the single peaked structure in
CeCoIn$_{5}$ to a monotonic decrease for the other systems. Vertical dashed
lines show the pressure where a maximum in $T_{c}(P)$ has been observed: 2.4
GPa for CeRhIn$_{5}$,\cite{Hegger00,Mito01} 1.4 GPa for CeCoIn$_{5}%
$,\cite{Sidorov02,Sparn02} 2.9 GPa for CeIrIn$_{5}$,\cite{Muramatsu03} and 2.4
GPa for Ce$_{2}$RhIn$_{8}$.\cite{Nicklas03}

\begin{figure}
[ptb]
\begin{center}
\includegraphics[
height=2.9162in,
width=3.5751in
]%
{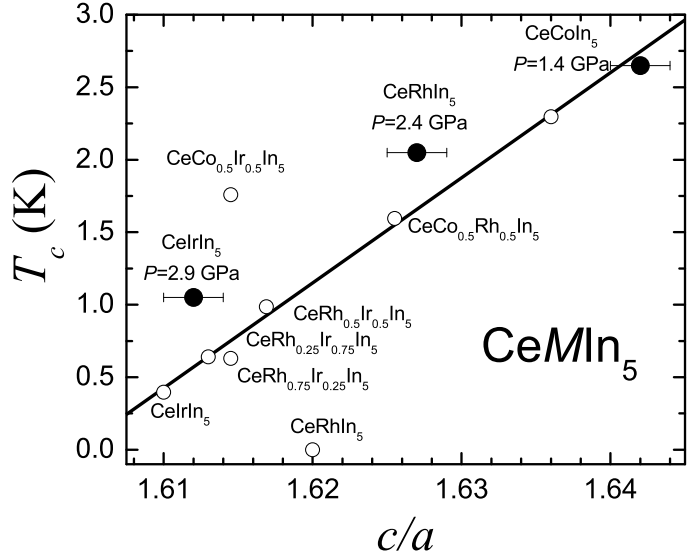}%
\caption{The ambient pressure values of the superconducting transition
temperature versus the room temperature value of $c/a$ (open circles) for
various Ce$M$In$_{5}$ compounds. Also shown (solid circles) are the values of
$c/a$ determined at room temperature at the pressure $P_{\max}$ where
$T_{c}(P)$ displays a maximum. The line is a least squares fit to the ambient
pressure values.}%
\label{tcvscoa}%
\end{center}
\end{figure}

As mentioned, a\ strong correlation between the ambient pressure $c/a$ ratio
and $T_{c}$ in the Ce$M$In$_{5}$ compounds has been observed (increasing $c/a$
increases $T_{c}$).\cite{Pagliuso02} This can be seen in Fig. \ref{tcvscoa}
that is adapted from Pagliuso \textit{et al}.\cite{Pagliuso02} (Note that we
have corrected a labeling error found in Pagliuso \textit{et al}%
.\cite{Pagliuso02} where two systems are labeled as CeCo$_{0.5}$Ir$_{0.5}%
$In$_{5}.$) However, some discrepancies exist, namely magnetic systems like
CeRhIn$_{5}$ whose $c/a$ ratio of 1.62 would lead one to erroneously conclude
that superconductivity near 1.0 K\ should be observed, rather than the
experimentally observed AF\ order at 3.8 K. The reason for this discrepancy
can be seen if one considers theoretical treatments of magnetically mediated
superconductivity.\cite{Monthoux01} Calculations show that superconductivity
occurs at a QCP\ where long range magnetic order is suppressed and the
infinite range magnetic correlations give way to short range magnetic
correlations that are responsible for the superconductivity. Recent work has
shown a similar sort of behavior when a system is near a valence instability
and critical density fluctuations give rise to
superconductivity.\cite{Monthoux04} In either of theses scenarios, one then
finds $T_{c}(P)$ behavior that displays the experimentally observed inverse
parabolic behavior with the maximum value of $T_{c}$ becoming larger as
correlations become more 2D in character. Slight deviations from the inverse
parabolic behavior observed in CeRhIn$_{5}$ on the high pressure
side\cite{Muramatsu01} may be indicative of density fluctuations or a "hidden"
3D magnetically ordered state.\cite{Nicklas04} In the magnetic fluctuation
scenario, the maximum value of $T_{c}(P)$ is found at a pressure $P_{\max}$
and depends on the spin fluctuation temperature $T_{sf}$ and the
dimensionality of the magnetic interactions. The maximum possible values of
$T_{c}$ will occur for more 2D systems with the highest possible value of
$T_{sf}$. This leads to the natural conclusion that the correct quantities to
plot are not the ambient pressure ones, but rather the value of $T_{c}$ at
$P_{\max}$ and the corresponding value of $c/a$. Note that while one should
use the structural information near $T_{c}$, we have shown that the $c/a$
versus $P$ behavior is similar at room temperature and near $T_{c}$ leading to
the conclusion that the room temperature lattice constants can be used for our
analysis.\cite{Kumar04} This has been done in Fig. \ref{tcvscoa} where the
filled circles correspond to the $c/a$ ratios from the current study where
$T_{c}$ reaches its maximum value at $P_{\max}$ taken from the
literature.\cite{Hegger00,Mito01,Sidorov02,Sparn02,Muramatsu03} As can be
seen, CeRhIn$_{5}$ now fits in with the rest of the data quite well. Also,
CeIrIn$_{5}$ and CeCoIn$_{5}$ both have their values of $T_{c}$ and $c/a$
enhanced from their ambient pressure values. Note that all of the points from
the current study lie on or above the line. These results are consistent with
theory and it would be of great interest to measure more values of the maximum
$T_{c}$ as a function of $c/a$ at that pressure to look for universal
behavior.%

To conclude that the dependence of $T_{c}$ on $c/a$ in Fig. \ref{tcvscoa} is
due mainly to dimensionality, it is necessary to prove that $T_{sf}$ does not
change drastically for the various compounds. To estimate $T_{sf}$, we have
used the tight binding approximation of Harrison to calculate the
hybridization $V_{pf}$ between the Ce (or Pu) $f$-electrons and In (or Ga)
$p$-electrons and $V_{df}$ between the Ce $f$-electrons and $M$ atom
$d$-electrons. As $T_{sf}\propto\exp(-1/V^{2}),$ the hybridization can be
directly linked to $T_{sf}$. It can be shown that the $pf$ and $df$
hybridization are given by
\begin{align}
V_{pf} &  =\eta_{pf}\frac{\hbar^{2}}{m_{e}}\frac{\sqrt{r_{p}r_{f}^{5}}}{d^{5}%
},\\
V_{df} &  =\eta_{df}\frac{\hbar^{2}}{m_{e}}\frac{\sqrt{r_{d}^{3}r_{f}^{5}}%
}{d^{6}},\nonumber
\end{align}
where $\eta$ is a constant (for $\sigma$ bonds, $\eta_{pf}=10\sqrt{21}%
/\pi,\eta_{df}=450\sqrt{35}/\pi$); $m_{e}$ is the mass of an electron;
$r_{p},r_{d}$ and $r_{f}$ are tabulated electron wavefunction radii for a
particular atom; and $d$ is the distance between the atoms in
question.\cite{Harrison80,Harrison83,Straub85,Harrison87} We tabulate ambient
pressure values along with values at the pressure where $T_{c}$ reaches its
maximum value $P_{\max}$ of both the $fp$ $(V_{fp})$ and the $df$ $(V_{df})$
hybridization, summing over all nearest neighbors, in Table \ref{hybrid}.
\begin{table}[ptb]
\caption{Calculated $fp$ $(V_{fp})$ and the $df$ $(V_{df})$ hybridization in
eV as described in text. Values are given at ambient pressure and the pressure
where $T_{c}$ displays a maximum $P_{\max}.$ Necessary structrual paramaters
for PuCoGa$_{5}$ are taken from Wastin \textit{et al}.\cite{Wastin03} and for
Ce$_{2}$CoIn$_{8}$ from Kalychak \textit{et al}.\cite{Kalychak89}}%
\narrowtext
\par%
\begin{tabular}
[c]{|c|c|c|c|c|c|}\hline
System & $V_{df}(0)$ & $V_{pf}(0)$ & $P_{\max}$ (GPa) & $V_{df}(P_{\max})$ &
$V_{pf}(P_{\max})$\\\hline
CeRhIn$_{5}$ & 0.572 & 2.030 & 2.4 & 0.607 & 2.136\\\hline
CeIrIn$_{5}$ & 0.627 & 2.031 & 2.9 & 0.665 & 2.135\\\hline
CeCoIn$_{5}$ & 0.307 & 2.066 & 1.4 & 0.317 & 2.130\\\hline
Ce$_{2}$RhIn$_{8}$ & 0.272 & 1.977 & 2.4 & 0.292 & 2.086\\\hline
Ce$_{2}$IrIn$_{8}$ & 0.297 & 1.993 & - & - & -\\\hline
Ce$_{2}$CoIn$_{8}$ & 0.147 & 2.018 & - & - & -\\\hline
PuCoGa$_{5}$ & 0.955 & 5.229 & - & - & -\\\hline
\end{tabular}
\label{hybrid}%
\end{table}Note that though we have done the calculation only for $\sigma$
bonds, the inclusion of bonding with higher $m$ quantum numbers will simply
multiply the final result by a constant (that should approximately be the same
for all members of an isostructural series). From Table \ref{hybrid}, it is
evident that $V_{pf}>V_{df}$ for all of the compounds. This is consistent with
the electronic structure calculations of Maehira \textit{et al}. that consider
the $fp$ hybridization only and get good agreement to measured Fermi
surfaces.\cite{Maehira03} This dominance of the $fp$ hybridization also gives
a natural explanation to some facts regarding the robustness of
superconductivity. For $M$ site substitution, superconductivity is robust and
exists for numerous Ce$M$In$_{5}$ compositions.\cite{Pagliuso01,Pagliuso02}
Substitution of Sn for In, however, has been shown to rapidly suppress
superconductivity in CeCo(In$_{1-x}$Sn$_{x}$)$_{5}$.\cite{Bauer04} These
results show that the $M$ atom serves mainly to affect the spacing between the
Ce and In atoms that determine the hybridization, and the sensitivity to Sn
substitution shows that disorder of the Ce-In strongly perturbs the $pf$
interactions leading to superconductivity.

\begin{figure}
[ptb]
\begin{center}
\includegraphics[
height=2.7691in,
width=3.9764in
]%
{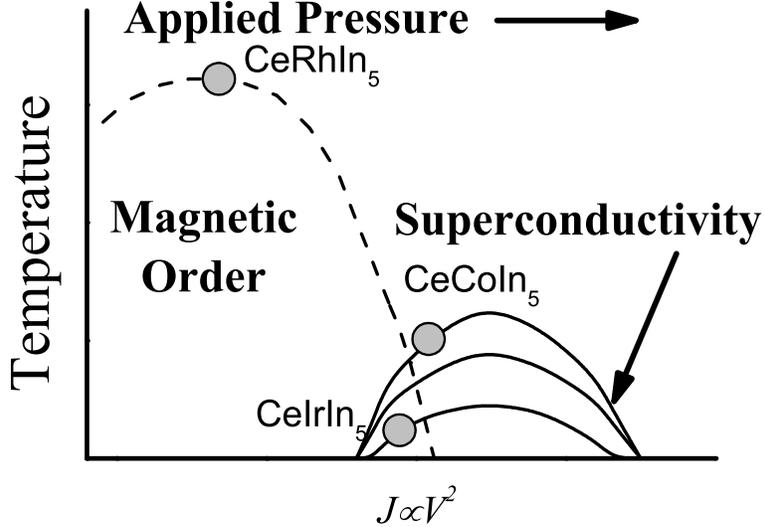}%
\caption{Schematic phase diagram for the Ce$M$In$_{5}$ compounds showing the
competition between magnetic order and superconductivity. For small values of
the hybridization $V^{2},$ the magnetically ordered state (dashed line) is
favored. As pressure is applied, systems move to the right in the diagram and
the magnetically ordered state gives away to superconductivity (solid lines).
The approximate ambient pressure position is shown for various Ce$M$In$_{5}$
materials. The superconducting curve for CeRhIn$_{5}$ lies between the
CeIrIn$_{5}$ and CeCoIn$_{5}$ curves.}%
\label{doniach}%
\end{center}
\end{figure}

For the Ce$M$In$_{5}$ series, the $V_{pf}$ values increase in the order
Rh$\rightarrow$Ir$\rightarrow$Co. One expects the important parameter
describing the magnetic interaction to be the magnetic coupling $J\propto
V^{2}.$ This is consistent with a Doniach model \cite{Doniach77,Doniach77_2}
of the competition between the nonmagnetic Kondo state and the magnetic
RKKY\ state shown schematically if Fig. \ref{doniach} which qualitatively
captures the pressure dependent behavior in Ce$M$In$_{5}$ compounds.%
After a system has reached its maximum magnetic ordering temperature, the
magnetic order is rapidly suppressed and the system moves toward a QCP. This
type of behavior has been seen in numerous Ce
compounds.\cite{Thompson94_2,Cornelius94,Cornelius97_2} Near the QCP, many
different behaviors can be observed. For the Ce$M$In$_{5}$ compounds,
superconductivity with a characteristic inverse parabolic shape is observed.
As shown by the dotted line, magnetic order may or may not coexist in regions
with superconductivity. In Fig. \ref{doniach}, the compounds were placed from
left to right in order of increasing $V_{pf}$. The location was chosen to
agree with the measured behavior of all three compounds. Namely, CeRhIn$_{5}$
is an antiferromagnet at ambient pressure while CeIrIn$_{5}$ and CeCoIn$_{5}$
are ambient pressure superconductors, and all three display a maximum in
$T_{c}$ as a function of pressure. The inverse parabolic shape of $T_{c}$ is
consistent with the behavior expected for magnetically mediated
superconductivity, where the height of the maximum depends on the
hybridization and the dimensionality.\cite{Monthoux01} The larger maximum
value of $T_{c}$ as a function of pressure for CeCoIn$_{5}$ with larger $c/a$
(and hence more 2D character) relative to CeIrIn$_{5}$ then follows naturally.
From Fig. \ref{doniach}, one would expect that the pressure to reach the
maximum in $T_{c}$ would increases in order Rh$\rightarrow$Ir$\rightarrow$Co.
Surprisingly, both Rh and Ir display the maximum at about the same pressure of
2.4 GPa. This can be explained, however, by noting that CeIrIn$_{5}$ has the
larger bulk modulus so that while the pressure is the same, the volume change
is considerably less. A more reasonable variable to use than pressure would be
the hybridization $V.$ From Table \ref{hybrid}, the value for the
hybridization at the pressure $P_{\max}$ where $T_{c}$ reaches its maximum
value is nearly identical for all three Ce$M$In$_{5}$ compounds. This gives
strong support for the magnetically mediated superconductivity scenario as one
would expect that the maximum value of $T_{c}$ would occur for approximately
the same value of $V$ and variations in $T_{c}$ would then be attributed to
differences in dimensionality. We note that the values of $V_{pf}$ for the
Ce$_{2}M$In$_{8}$ compounds is very similar to the Ce$M$In$_{5}$ compounds and
the progression of increasing $V_{pf}$ being Rh$\rightarrow$Ir$\rightarrow$Co;
this is consistent with the progression of ground states from magnetic order
(Rh) to heavy fermion (Ir) to superconductivity (Co) in the Ce$_{2}M$In$_{8}$
series. This is in line with the experimental finding of very similar
electronic specific heat coefficients $\gamma\propto1/T_{sf}$ $\propto
\exp(1/V^{2}$). \cite{Cornelius01,Moreno02,Thompson03} Also, in a scenario of
magnetically mediated superconductivity, the most obvious route to higher
$T_{c}$ values would be to raise the value of $T_{sf}$ by switching to
actinide compounds with larger $r_{f}$ values, and hence hybridization
relative to rare earths. The affect of moving to the actinides is seen in
PuCoGa$_{5}$ that has $V_{pf}$ $\sim2.6$ times larger than the corresponding
Ce compounds.

\begin{figure}
[ptbptb]
\begin{center}
\label{fitparams}%
\includegraphics[
height=2.9162in,
width=3.7567in
]%
{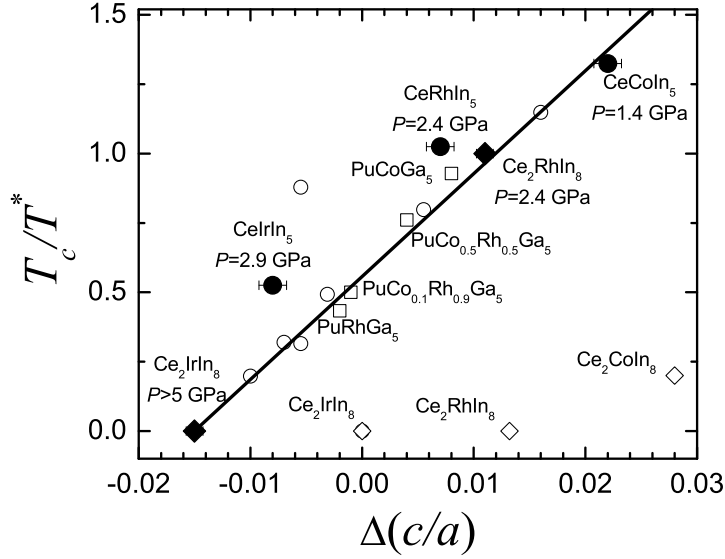}%
\caption{The ambient pressure values of $T_{c}/T^{\ast}$ versus the room
temperature value of $\Delta(c/a)$ (open symbols) for various Ho$_{n}%
$CoGa$_{2n+3}$ based compounds; Ce$M$In$_{5}$ (circles), Ce$_{2}M$In$_{8}$
(diamonds) and Pu$M$Ga$_{5}$ (squares) are all shown. Also shown (solid
symbols) are the values of $\Delta(c/a)$ determined at room temperature at the
pressure $P_{\max}$ where $T_{c}(P)$ displays a maximum. The straight line is
the same as that shown in Fig. 4.}%
\label{coanorm}%
\end{center}
\end{figure}

Recently, Pu based superconductivity was observed for the first time in
PuCoGa$_{5}$ above 18 K, an order of magnitude larger than the Ce compounds
that also have the HoCoGa$_{5}$ structure.\cite{Sarrao02} It was subsequently
shown by Wastin \textit{et al}. that a similar universal linear behavior of
$T_{c}$ versus $c/a$ is observed in Pu$M$Ga$_{5}$ compound with nearly the
same logarithmic slope as the Ce$M$In$_{5}$ compounds.\cite{Wastin03,Wastin04}
While this may at first seem a surprising result, in fact it follows straight
from the theoretical conclusions that $T_{c}$ should scale as a characteristic
temperature $T^{\ast}\propto$ $T_{sf}$ . That the value $T_{c}$ is an order of
magnitude larger in Pu based compared to Ce based compounds then is a
consequence of a value of $T_{sf}$ that is an order of magnitude larger in Pu
compounds. This estimate is reasonable in light of the previous discussion
showing a significantly larger value of $V_{pf}$ in the Pu compounds
remembering that $T_{sf}\propto\exp(-1/V^{2}$), and also because the
electronic specific heat coefficient $\gamma$ is an order of magnitude smaller
in Pu compounds relative to Ce compounds and $T_{sf}\propto1/\gamma
.$\cite{Sarrao02} We also note that the Ce$_{2}M$In$_{8}$ compounds at ambient
pressure do not seem to not follow the linear $T_{c}$ versus $c/a$ behavior as
only Ce$_{2}$CoIn$_{8}$ displays superconductivity at ambient pressure.
However, Ce$_{2}$RhIn$_{8}$, like CeRhIn$_{5},$ magnetically orders at ambient
pressure but the application of pressure reveals superconductivity. To further
analyze these systems, we plot normalized values of $T_{c}$ versus $\Delta
c/a$ in Fig. \ref{coanorm}, where $T_{c}$ is normalized by $T^{\ast}$ and
$\Delta(c/a)$ is found by subtracting a value $(c/a)^{\ast}$. $T^{\ast}$ was
chosen as 2 K for Ce$M$In$_{8}$ and Ce$_{2}M$In$_{8}$ as it is approximately
$T_{sf}$ for CeCoIn$_{5}$,\cite{Nakatsuji02} and as discussed previously, we
don't expect much variation in $T_{sf}$ for these compounds. $T^{\ast}=20$ K
was used for Pu$M$Ga$_{5}$ as we expect an order of magnitude increase in
$T_{sf}$ for Pu compounds relative to Ce compounds. $(c/a)^{\ast}$ was chosen
in such a way to shift the curves on top of each other. The values of
$T^{\ast}$ and $(c/a)^{\ast}$ are given in Table \ref{fitparams}.
\begin{table}[ptb]
\caption{ Summary of the normalization values used to plot the data in Fig.
\ref{coanorm}. $T^{\ast}$ is a characteristic temperature that is related to
the spin fluctuation or Kondo temperature. $(c/a)^{\ast}$ is chosen as
described in text.}%
\label{fitparams}
\narrowtext%
\begin{tabular}
[c]{|c|c|c|}\hline
System & $T^{\ast}(K)$ & $(c/a)^{\ast}$\\\hline
Ce$M$In$_{5}$ & $2.0$ & $1.620$\\\hline
Pu$M$Ga$_{5}$ & $20$ & $1.596$\\\hline
Ce$_{2}M$In$_{8}$ & $2.0$ & $2.610$\\\hline
\end{tabular}
\end{table}The normalized values are plotted in Fig. \ref{coanorm}. The
universality is readily apparent with all of the pressure points lying on or
above the straight line. That the points lie on or above the line for the
ambient pressure points is likely due to higher values of $T_{sf}$ for the
optimal pressure data relative to ambient pressure data rendering the
assumption of a single $T^{\ast}$ value to normalize all data tenuous. The
ambient pressure "misplacement" of Ce$_{2}$RhIn$_{8}$ (AF\ order at ambient
pressure) now can be explained by the pressure induced superconductivity and
the universal line now goes through the high pressure Ce$_{2}M$In$_{8}$ data.
While Ce$_{2}$IrIn$_{8}$ does not display superconductivity, the value of
$c/a$ reaches a nearly constant value above 5 GPa and we have plotted a point
assuming $T_{c}=0$ at high pressure. This assumption gains validity as these
results would predict that superconductivity will not be seen in Ce$_{2}%
$IrIn$_{8}$ under pressure as $\Delta(c/a)$ falls below the x-intercept of the
$T_{c}/T^{\ast}$ versus $\Delta(c/a)$ line. Also, Ce$_{2}$CoIn$_{8}$ should
see a dramatic enhancement of $T_{c}$ under pressure; if $c/a$ doesn't change
as a function of pressure, this estimate for the maximum in $T_{c}$ would be
around 3 K which is slightly larger than what is seen in CeCoIn$_{5}$ under
pressure.

\section{Conclusions}

We have studied the elastic properties of Ce$_{n}M$In$_{2n+3}$ ($M$=Rh, Ir and
Co) with $n=1$ or 2 under hydrostatic and quasihydrostatic pressures up to 15
GPa using x-ray diffraction. The addition of $M$In$_{2}$ layers to the parent
CeIn$_{3}$ compound is found to stiffen the lattice. By plotting the maximum
values of the superconducting transition temperature $T_{c}$ versus $c/a,$ we
are able to expand upon the proposed linear relationship between the
quantities by Pagliuso \textit{et al}.\cite{Pagliuso02} We have also found
that the dominant hybridization is between the Ce (or Pu) $f-$electrons and In
(or Ga) $p-$electrons $V_{pf}$. Also, the value of $V_{pf}$ where $T_{c}$
reaches its maximum is nearly identical for all three Ce$M$In$_{5}$ compounds.
These results explain the lack of superconductivity in Ce$_{2}$IrIn$_{8}$ and
predict that $T_{c}$ should increase dramatically in Ce$_{2}$CoIn$_{8}$ at
high pressure. Comparing the results to Pu-based superconductors shows a
universal $T_{c}$ versus $c/a$ behavior when these quantities are normalized
by appropriate quantities consistent with what is expected of magnetically
mediated superconductivity.

\begin{acknowledgments}
Work at UNLV is supported by DOE EPSCoR-State/National Laboratory Partnership
Award DE-FG02-00ER45835. Work at LANL is performed under the auspices of the
U.S. Department of Energy. HPCAT is a collaboration among the UNLV High
Pressure Science and Engineering Center, the Lawrence Livermore National
Laboratory, the Geophysical Laboratory of the Carnegie Institution of
Washington, and the University of Hawaii at Manoa. The UNLV High Pressure
Science and Engineering Center was supported by the U.S. Department of Energy,
National Nuclear Security Administration, under Cooperative Agreement
DE-FC08-01NV14049. Use of the Advanced Photon Source was supported by the U.
S. Department of Energy, Office of Science, Office of Basic Energy Sciences,
under Contract No. W-31-109-Eng-38.
\end{acknowledgments}

\bibliographystyle{apsrev}
\bibliography{acompat,cornel}

\end{document}